Observation of Longitudinal Spin Seebeck Effect with Various Transition Metal Films

M. Ishida[1*], A. Kirihara[1], H. Someya[1], K. Uchida[2,3], S. Kohmoto[1], E. Saitoh[2,4,5,6], and T. Murakami[1]

[1] Smart Energy Research Laboratories, NEC Corp, Tsukuba 305-8501, Japan

[2] Institute for Materials Research, Tohoku University, Sendai 980-8577, Japan

[3] PRESTO, Japan Science and Technology Agency, Saitama 332-0012, Japan

[4] Advanced Science Research Center, Japan Atomic Energy Agency, Tokai 319-1195, Japan

[5] CREST, Japan Science and Technology Agency, Tokyo 102-0075, Japan

[6] WPI Advanced Institute for Materials Research, Tohoku University, Sendai 980-8577, Japan

We evaluated the thermoelectric properties of longitudinal spin Seebeck devices by using ten different transition metals (TMs). Both the intensity and sign of spin Seebeck coefficients were noticeably dependent on the degree of the inverse spin Hall effect and the resistivity of each TM film. Spin dependent behaviors were also observed under ferromagnetic resonance. These results indicate that the output of the spin Seebeck devices originates in the spin current.

Thermoelectric generation (TEG) based on the spin Seebeck effect (SSE) has received a lot of research and industrial interest because of its ability to generate spin and electrical currents separately [1-3]. In spin TEG using the longitudinal SSE (LSSE), a heat flow applied to a ferro/ferrimagnetic insulator (FI) accompanies the flow of the spin angular momentum (spin current) along the heat flow. The spin current is then injected into a paramagnetic transition metal (TM) attached to the FI by interference between the dynamics of localized spins in the FI and conduction electrons in the TM. The spin current produced in the TM, which is described as a pure spin current, can be converted into an electrical current by means of an inverse spin Hall effect (ISHE) in the TM.

The maximum power output ($P_{MAX} = V^2/4R$ [W/m$^2$]) of the spin TEG depends on the following expression [2-4].

$$P_{MAX} \propto (\Delta T_{me} g_r \theta_{SH})^2 \rho_N V_N, \tag{1}$$

where $\Delta T_{me}$ is the effective temperature difference between magnons in the FI and electrons in the TM at the FI/TM interface, $g_r$ is the real part of spin mixing conductance that represents the efficiency of spin pumping at the interface, $\theta_{SH}$ is the spin Hall angle that determines the efficiency of spin current-electrical current conversion due to ISHE, and $\rho_N$ and $V_N$ are the electrical resistivity and volume of TM. $\Delta T_{me}$ can be increased by modifying the material, quality, and thickness of FI [3]. $g_r$ is related to characteristics of the interface such as the spin density [5] and also the integrity of fabrication processes [6].

For TMs undertaking $\theta_{SH}$, platinum has been mainly adopted because it has the largest $\theta_{SH}$ among the TMs [7-9]. To explore a larger $\theta_{SH}$ and more cost-efficient materials, the spin Hall effect (SHE) and ISHE of a variety of TMs, alloys, and an oxide have been studied more extensively in spin pumping experiments and other spintronic applications [10-17]. In spincalorinics applications including SSE, a few TMs have been investigated [18-20].

In this study, we verified ten different TMs to collect spin-TEG outputs. A comparison between the spin-TEG properties and previous theoretical/experimental results on the SHE/ISHE of these metals would be of assistance in understanding such spin current-related phenomena collectively and also in improving device performances.

The spin-TEG devices evaluated in this study were fabricated with the same methods described in the previous study [3]. 120-nm-thick bismuth substituted yttrium iron garnet (BiY$_2$Fe$_5$O$_{12}$, BYIG) films were grown with the metal-organic decomposition (MOD) method. Substituted gadolinium gallium garnet ((GaGd)$_3$(ZrMgGa)$_5$O$_{12}$, SGGG) was used for the substrates. A BYIG/SGGG wafer was then cut into 8 × 2 mm slabs and 10-nm-thick TM films (Ti, Cr, Cu, Mo, Ag, Ta, W, Ir, Pt, and Au) were deposited on the surface of the BYIG by using a commercial magnetron sputter deposition system with the same sputtering conditions. The deposition time was changed only in order to adjust the thickness. In TE measurements, a temperature difference was applied over a 2 × 4 mm area by making two temperature-controlled blocks on both the Pt and SGGG faces contact. The open circuit voltage $V_{ISHE}$ that would appear in the TM film was measured with two contact probes separated by 6 mm. The entire system was set up in the hollow core of an electromagnet to apply a uniform magnetic field. The TE measurements for the Cu film were carried out right after the deposition to minimize the effect of oxidation [21].

Figure 1(a) shows the schematics of the device structures. To determine the effect of the spin current generated from BYIG and to exclude intrinsic effects in the Pt film, two types of devices, Pt/BYIG/SGGG and Pt/SGGG, were fabricated. $V_{ISHE}$ profiles from the two devices were taken while changing the temperature difference $\Delta T$ and external magnetic field $H$ along the z- and x-axes, respectively. As shown in Fig. (b), when a certain temperature gradient was applied, distinct $V_{ISHE}$ signals appeared as $H$ increased only in the case of Pt/BYIG/SGGG and saturated around $H^{sat} = \pm 300$ [Oe], which was the almost same as the saturation magnetic field of the 120-nm-thick BYIG. We defined the saturation voltage at $\Delta T$ as $V_{ISHE}^{sat}(\Delta T) = (V_{ISHE}(H_+^{sat}) - V_{ISHE}(H_-^{sat}))/2$. Figure 1(c) shows the $V_{ISHE}^{sat}(\Delta T)$ plots for the two devices. A spin-Seebeck coefficient $S_{SSE}$ of 1.64 (±0.05) µV/K for Pt/BYIG/SGGG and no distinct $S_{SSE}$ for Pt/SGGG were estimated from the slopes of the plot.

In addition to Pt, we fabricated nine different TM films (Ti, Cr, Mo, Ta, W, Cu, Ag, Ir, and Au) with a thickness of 10 (±1) nm on BYIG/SGGG slabs taken from one wafer. Figure 2 shows a summary of the $V_{ISHE}(H)$ profiles obtained at $\Delta T = 8$ K. A clear $V_{ISHE}$ was observed in the devices with Cr, Mo, Ta, W, Ir, and Au films.

In Fig. 3, $S_{SSE}$, $\rho_N$, and the power factor ($P.F.$ $[W/mK^2]$) were estimated. Pt, Au, and Ir/BYIG generated a large $S_{SSE}$ in the same sign as the Pt system. In contrast, Cr, Mo, Ta, and W/BYIG revealed large signals of $S_{SSE}$ with the opposite sign. No signal was observed in the TM/SGGG systems without BYIG. We found that, in comparison with the result of the Pt system, $S_{SSE}$ intensities and signs of the other TMs, especially in period 6, resulted in having a good consistency with the theoretical and experimental studies on SHEs [8,9,11]. This result indicates that the origin of $V_{ISHE}$ in the TMs was the thermal spin current injected from BYIG into the TMs, which is irrelevant to the anomalous Nernst effect induced by a static magnetic proximity [22]. Notably, W exhibits a large $S_{SSE}$ despite its electronic structure being far from Stoner ferromagnetic instability [23,24]. The observed large values of $|S_{SSE}|$ in the TM/BYIG structures could be attributed to the large spin-orbit interaction or high resistivity of TMs, as described in equation (1). Note that, with a variation of the TMs, modulation of the spin mixing conductance $g_r$ should also be expected to change $S_{SSE}$. In this result, however, no distinct dependence other than that of $\theta_{SH}$ and $\rho_N$ was revealed. Further studies are needed to clarify the modulation of $g_r$ experimentally. With respect to $P.F.$, Pt/BYIG had the largest value (5.8 × $10^{-6}$ [W/mK$^2$]). The advantage of Au was enhanced up to 8.4 × $10^{-7}$ [W/mK$^2$] by its lower resistivity. In comparison, for Ta and W, the high resistivity $\beta$-phase, which is particularly observed in thin films approximately less than 10 nm, was reported to enhance $\theta_{SH}$ [16,17]. The results in Fig. 3 indicated that the Ta and W films were a mixture of high and low resistivity phases and that $S_{SSE}$ and $P.F.$ would be increased further by the use of the pure $\beta$-phase.

To confirm this spin-related phenomenon, we performed simultaneous measurements of $V_{ISHE}$ and microwave absorption under ferromagnetic resonance (FMR). Figure 4 (a) shows $V_{ISHE}(H)$ obtained from Pt/BYIG and W/BYIG under microwave irradiation at a frequency of 9.44 GHz with an input power of 10 mW. The two voltage signals were observed with opposite signs. The result was in good agreement with the results of the spin-TEG measurements. The peak intensities were estimated by fitting with the Lorentzian function to be 10.2 and -5.7 μV for Pt/BYIG and W/BYIG, respectively.

The FMR absorption spectrum $dI(H)/dH$ and differentials of the voltage signal $dV_{ISHE}(H)/dH$ were then compared. In Fig. 4(b), the two spectra obtained from Pt/BYIG were perceived to correspond very well and were reproducible with the simple Lorentzian function. The same relation was also observed in W/BYIG (Fig. 4(c)), whereas $dV_{ISHE}/dH$ was turned over in accordance with the sign of the voltage. These results clearly indicate that the voltage signal measured in this experiment could be recognized as the result of ISHE induced by spin pumping [25].

In summary, we investigated the $S_{SSE}$, $\rho_N$, and $P.F.$ of spin-TEG devices with various TM films. We concluded that $S_{SSE}$ depends on the degree of ISHE and also the $\rho_N$ of each TM film. The voltage measurements under the FMR condition also supported the presence of the spin-related phenomenon. The $P.F.$ of Au/BYIG enhanced by the lower resistivity was next to the Pt system. W also showed a large $P.F.$ due to the enhanced $S_{SSE}$ of the high resistivity $\beta$-W.

The authors thank Y. Kajiwara for his assistance in the FMR measurements. This work was supported by PRESTO-JST's "Phase Interfaces for Highly Efficient Energy Utilization," CREST-JST's "Creation of Nanosystems with Novel Functions through Process Integration," a Grant-in-Aid for Research Activity Start-up (24860003) from MEXT, Japan, a Grant-in-Aid for Scientific Research (A) (24244051) from MEXT, Japan, LC-IMR of Tohoku University, the Murata Science Foundation, the Mazda Foundation, and the Sumitomo Foundation.

Figure captions

Fig. 1: (a) Schematics of two spin-TEG devices (Pt/BYIG/SGGG and Pt/BYIG). Directions of temperature gradient $\Delta T$, external magnetic field H, and electric filed produced by ISHE $E_{ISHE}$ are defined. (b) $V_{ISHE}(H)$ profiles at $\Delta T$ = -8, -4, 0, 4, and 8 [K] and (c) $V_{ISHE}^{sat}(\Delta T)$ plots for the two spin-TEG devices.

Fig. 2: $V_{ISHE}(H)$ obtained from TM/BYIG/SGGG (TM = Ti, Cr, Mo, Ta, W, Cu, Ag, Ir, Pt, and Au) at $\Delta T = +8$ [K].

Fig. 3: Spin-Seebeck coefficient $S_{SSE}$, resistivity $\rho_N$ of TM films, and power factor ($P.F.$) of spin-TEG devices with TM/BYIG/SGGG (TM = Ti, Cr, Mo, Ta, W, Cu, Ag, Ir, Pt, and Au). The $\rho_N$ plot was a result of the four-probe measurement, and the resistance correction factor was also considered. For $P.F.$, however, we used two probe resistances from a practical point of view. The relation between $P.F.$ and maximum output power was $P.F.= S_{SSE}^2/\rho_N = P_{MAX} w d_N/l\Delta T^2$, where $l, w,$ and $d_N$ are the length, width, and thickness of the TM film.

Fig. 4: (a) External magnetic field dependence of $V_{ISHE}$ for Pt/BYIG/SGGG and W/BYIG/SGGG under microwave excitation near the ferromagnetic resonance (FMR). (b), (c) Comparisons between FMR absorption spectra $dI(H)/dH$ and differentials of the voltage signal $dV_{ISHE}(H)/dH$.

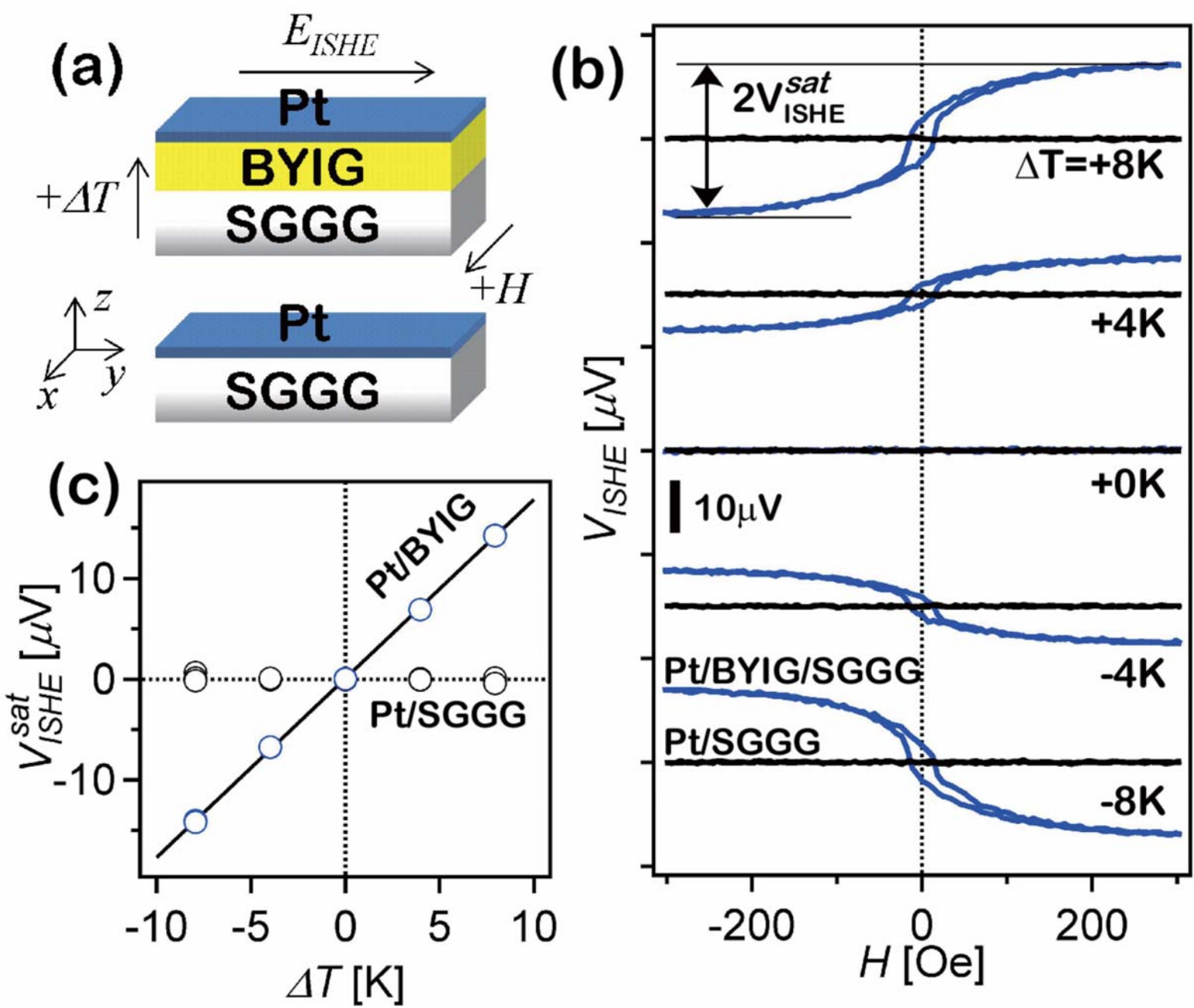

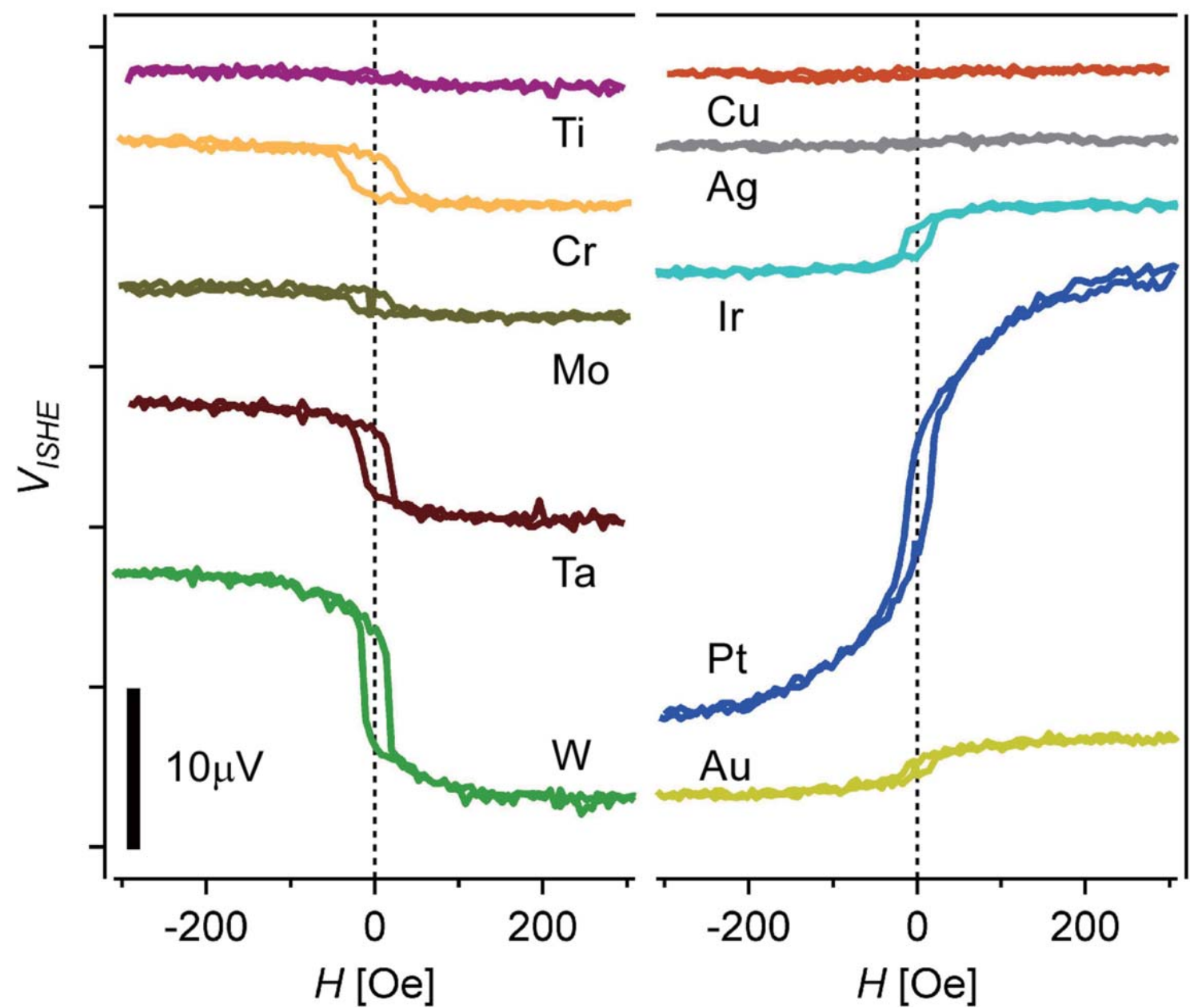

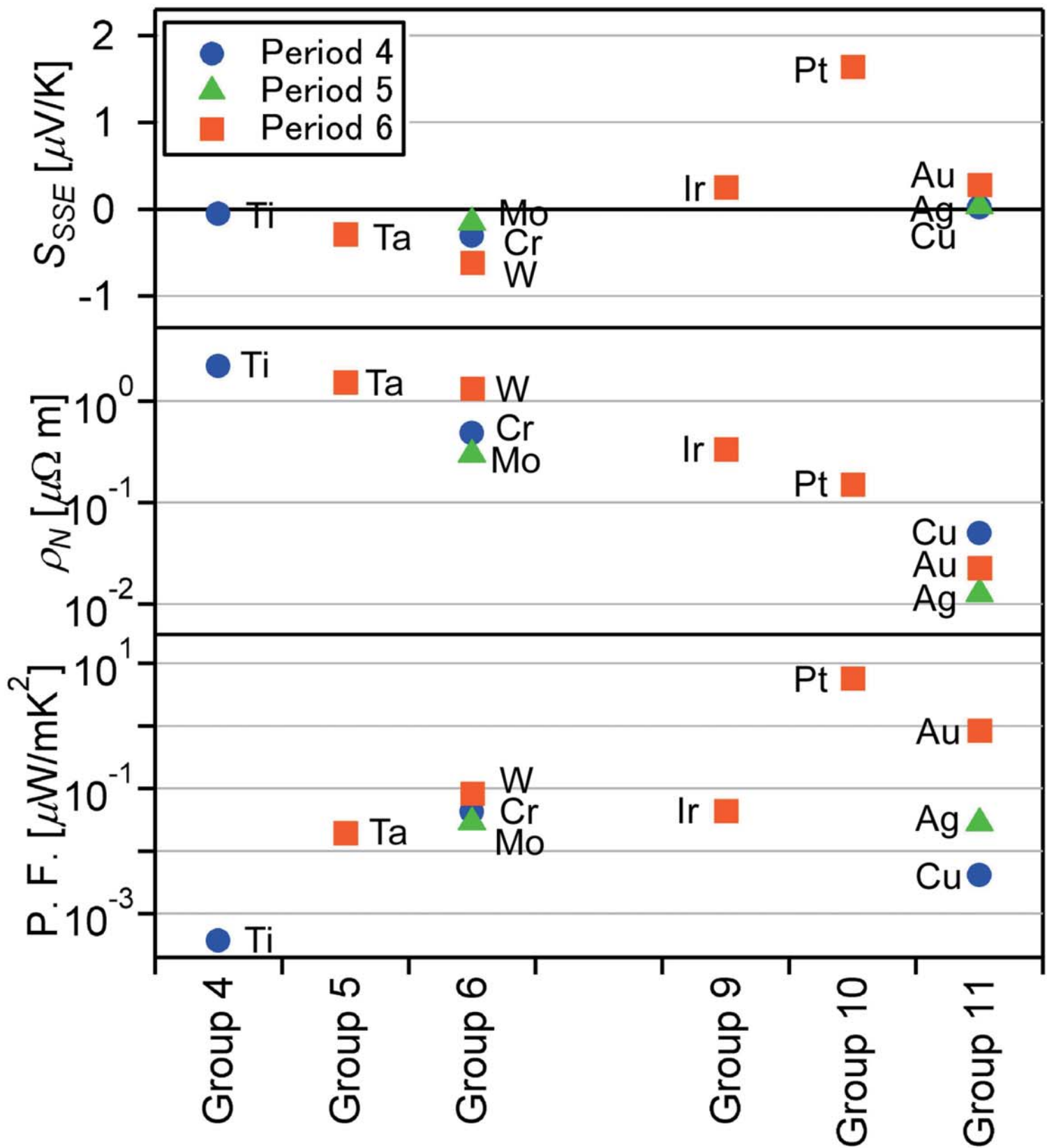

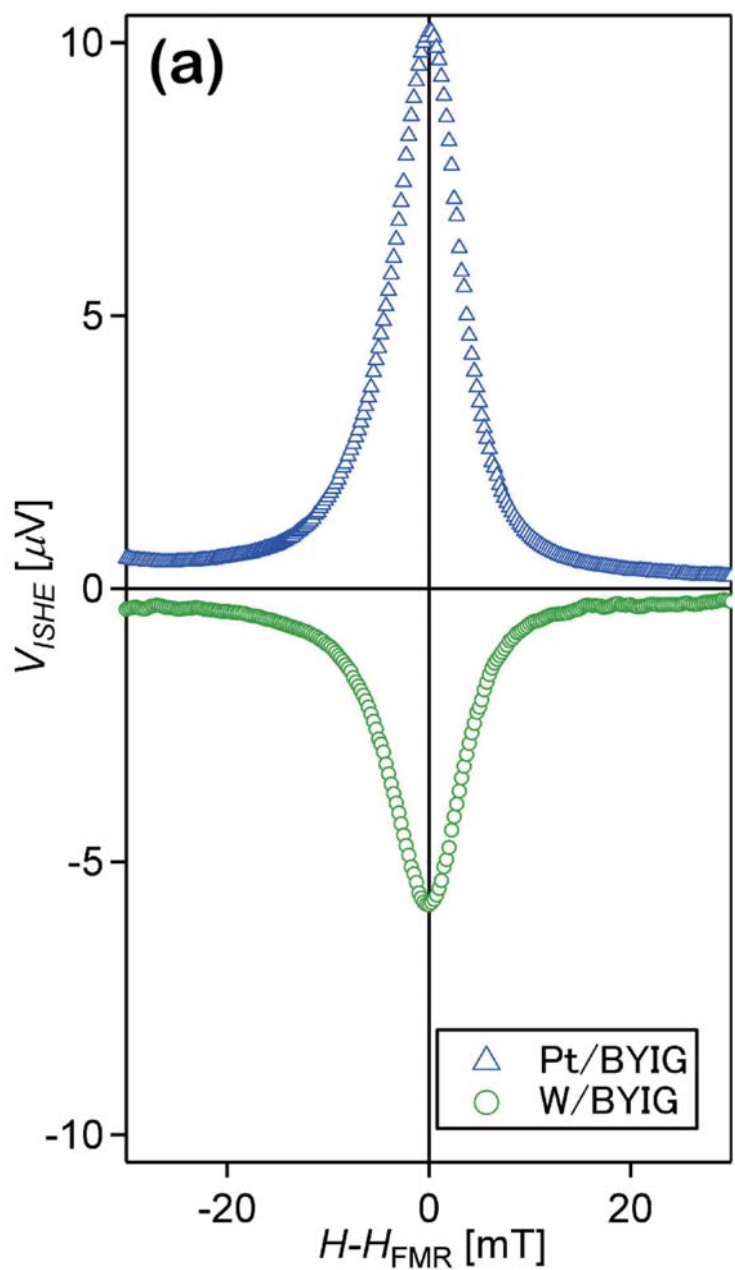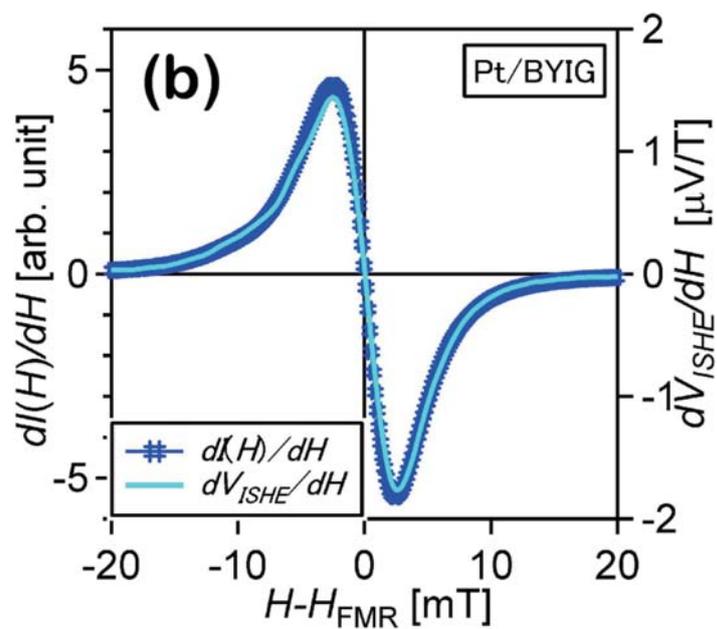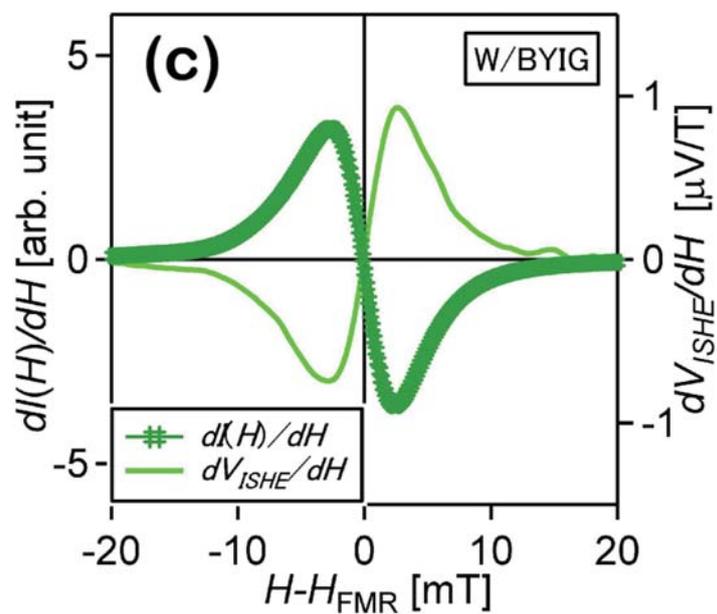